# Music of the spheres: sound emitted by the bubbles in liquid helium

P. B. Lerner[1]

Abstract

A large linewidth of electronic transition of an electron trapped in a bubble ("bubblonium") possesses natural, or radiative, and inhomogeneous components. The latter mechanism of the line broadening requires dissipation. Dissipation of optical-frequency quantum into a liquid helium leads to an emission of an acoustic wave. The possibilities of detection of this acoustic signal are discussed in this paper. Finally, the low temperature analogues with several hard-to-observe astrophysical phenomena are discussed.

1. **Introduction**

An electron trapped in a liquid helium bubble displays many properties of atomic and molecular emission, which was referred as "bubblonium" by this author [1] for its behavior. In particular, similar to the linewidth of an atom inside a gas, the bubblonium linewidth possesses radiative (homogenous) and non-radiative (inhomogeneous) components. While the radiative component is absolutely similar to the natural linewidth of an atomic transition, non-radiative component requires dissipation of energy into the ambience. [2]

The mechanisms of dissipation are very limited in the liquid He in comparison to conventional solid states. One of them is the emission of a number of acoustic modes. [3] I argue that the emission of acoustic waves can be detected and used to provide richness of information about bubbles themselves, as well as with their interaction with the environment.

2. **The estimates of the sound propagation**

A typical size of the bubble in the liquid helium kept in equilibrium by quantum pressure of a trapped electron is 18-19 Å in a 1S state and 22-24 Å in a 1P state. [4] From the general acoustics considerations we assume that the most efficient emission of an oscillating sphere in incompressible liquid will be radiated at a wavelength comparable to the radius of the sphere. [5] This assumption will be investigated quantitatively in subsequent sections.

---

[1] Retired.

Estimated modal frequency of the sound emission is thus determined by the following consideration:

$$\frac{\omega_m \cdot r_0}{v_s} \sim 1 \qquad (1)$$

or

$$\omega_m \sim \frac{v_s}{r_0} \qquad (2)$$

where $v_s$ is a sound velocity and $r_0$ is the radius of the bubble. Approximate estimate according to the to the equation (1) for a typical sound velocity in a superfluid $He^4$ at T=2 K° provides a value for a typical frequency of the emitted sound as $\omega_m \sim 10^{11}$ sec$^{-1}$ for the first sound and are $\omega_m \sim 10^{10}$ sec$^{-1}$ for the second. [6] The question of attenuation of the sound in the liquid $He^4$ and $He^3$ has been extensively studied but for the much lower frequencies on the order of 100 MHz. [7, 8] In the measurements by J. P. Davis *et al*. [8], the zero sound with frequencies of 88-147 MHz exhibited attenuation in the range of several hundred cm$^{-1}$ with very little dependence on frequency below $T/T_c$ =0.25. Their methods allowed for detection of sound waves with dissipation up to 1000 cm$^{-1}$. If we assume, following [6-8] that for the first and the second sound, the attenuation for the high-frequency domain grows proportionally to $\omega^2$, the scaling of the lower frequency results of [7, 8] leads to the $\alpha \geq 10^2 \div 10^3$ cm$^{-1}$. If one decreases temperature closer to the attenuation minimum (several tens of millikelvin), the intensity of the sound will decrease in proportion to some power of temperature. It is unclear whether sound dissipating so fast can be observed experimentally. However, the experiments with neutron scattering positively identify propagating waves in liquid He with wave vectors of 0.25-4 Å$^{-1}$. [9] Another possibility would be to observe the emission of zero sound in the liquid $He^3$.

## 3. Quasiclassical theory of sound emission

The problem of the scattering of virtual phonons on the bubblonium "breathing" as a result of quantum fluctuations, or thermal noise, is complicated. Yet, one can use the phonon analogue of the "macroscopic quantum electrodynamics." [10] In it, the computation is performed using c-numbers instead of quantum mechanical operators to replace them in the end with the expectation values obtained from quantum-mechanical commutation relations. This method, despite its apparent crudeness, produces accurate, even exact results, which can be confirmed in many cases by direct computation using Feynman diagrams. [11]

First, let us, use the classical expression [12] for the power scattered by a sphere undergoing harmonic oscillations. Distribution of the velocities on the surface of the sphere is as follows:

$$v(\vartheta,\varphi) = -i\omega A \cos(2\vartheta)e^{-i\omega t} \quad \quad 3)$$

where A is the amplitude of the wave and $\cos(2\cdot\theta)$ is the double cosine of the polar angle in the spherical coordinates related to the bubble. The scattered power is given by the expression:

$$P = \frac{128\pi\rho \cdot v_s^3 |A|^2}{45[D_2'(k\hat{a})]^2} \to \begin{cases} \dfrac{128}{3645}\pi\rho v_s^3 k^8 \hat{a}^8 |A|^2, & ka \to 0, \\ \dfrac{128}{45}\pi\rho v_s^3 k^2 \hat{a}^2 |A|^2, & ka \to \infty \end{cases} \quad 4)$$

where $D_n(z)$ is the amplitude of the associated spherical function $h_n$ [14]
$$h_n = j_n + i\cdot n_n = -i\cdot D_n e^{i\delta_n} \quad \quad 5)$$

and

$$D'_2 = \sqrt{|h'_2(z)|^2}$$

The bubble size variable with a hat means that strictly speaking the size of the bubble is a quantum mechanical operator, which has no separate meaning in our quasiclassical approach and is understood exclusively in terms of its expectation values for the multipole configurations of the bubble.

Generally, this formulation makes no physical sense (though, applied to dipole radiation of isolated atoms it is *exact*) but the applicability of the Frank-Condon principle, or Born-Oppenheimer approximation to the optical relaxation of the bubble assures that this is a decent approximation.[1] It simply means that for the scattering of massless particles, such as photons and phonons, the oscillations of the bubble are decoupled from the scattering and that in most contexts (see, however, the Section 5), the phenomena such as Doppler shift of incoming radiation and bubble recoil can be neglected. Because of the Born-Oppenheimer approximation one can consider separately oscillations of the bubble and quantum-mechanical amplitude of the scattered phonons.

While expressions such as Equation (4) are not well-defined, the expectations of the multipole matrix elements of them have a definite meaning. For instance, in absolute similarity with the dipole radiation case described in every textbook of quantum mechanics [13], the expectation of the square of the bubble "size operator" is equal to:

$$< L = 0|\hat{a}^2|L = 2 > = \sum_{k\prime=0}^{\infty} < L = 0|\hat{a}|L = k' >< L = k'|\hat{a}|L = 2 > = |< L = 0|\hat{a}|L = 2 >|^2$$

because of the selection rules for multipole matrix elements. Spherical Bessel functions from an operator understood as a power series are also well defined because of quick (faster than the Laurent series for an exponential) convergence of the Bessel functions. [14]

To determine the amplitude of the phonon wave, we have to equate the phonon energy to the energy of the quantum fluctuations in the mode with the wave vector k corresponding to the frequency of oscillations. Similar method was used by us in [4]. The energy per mode is given by conventional relation:

$$w_k = \frac{\rho v_s^2 k^2 |A_k|^2}{2} \qquad 6)$$

where the amplitude is computed as to reproduce the Planck frequency distribution for the phonons (bozons). The zero-point amplitude of the sound oscillations in the liquid at a temperature T

$$A_{0,k} = \left(\frac{4\pi\hbar}{\rho v_s kV}\right)^{1/2} \qquad 7)$$

and

$$A_k^2 = A_{0k}^2 \times n(\mathbf{k})$$

where n(**k**) is a conventional Plank density-of-states function. Using the equation (7) for the amplitude we can compute the power of the scattered light per mode as

$$P_k = \frac{2\pi\rho v_s^3 |A_k|^2}{3 \cdot [D'_2(ka)]^2} \qquad 8)$$

and the total emitted power as

$$N_{tot} = 4\pi \int P_k dV \frac{k^2 dk}{(2\pi)^3} \qquad 9)$$

Simplification of the expressions above, we get

$$N_{tot} = \frac{(k_B T)^2}{3\pi\hbar} \times \langle L=0|\hat{I}_1|L=2\rangle \qquad 10)$$

where

$$\hat{I}_1 = \int \frac{xdx}{\left[D'_2\left(\frac{k_B T \hat{a}}{c\hbar}x\right)\right]^2 (e^x - 1)} \qquad 11)$$

The expression <L=0|I$_1$|L=2> is purely symbolic meaning that the element of the density matrix of the oscillating bubble connects the excited and the ground state with their respective angular momenta.

Equation (10) can be simplified (and made computable) in the limiting cases $k_B \cdot T \cdot a / v_s \cdot \hbar \geq 1$ and $k_B \cdot T \cdot a / v_s \cdot \hbar \leq 1$, though in practice, the characteristic phonon wavelength in helium at 1-2 K° is of the same order as the wavelength for the oscillatory mode of a bubble.

For instance in the "high temperature" limit, where the characteristic phonon wavelength is smaller than the size of the bubble oscillation, the total emitted power is:

$$N_{tot} \approx \frac{(k_B T)^4 <\varepsilon_2^2> a^2}{3\pi \cdot c^2 \hbar^3} I_2 \qquad 12)$$

where

$$I_2 = \int_0^{+\infty} \frac{x^4 dx}{e^x - 1} = 24 \cdot \zeta(5) \cong 24.89 \qquad 13)$$

and $\varepsilon_2$ is a parameter describing the deformation of the sphere into an ellipsoid. [1] A crude estimate of square of its expectation value is provided in Appendix A. One observes that the law of emitted sound power in this limit is the same as for the blackbody radiation. The formula (12) can be mnemonically expressed as a "blackbody radiation of sound from an object with the size of quadrupole moment of oscillations."

This law is non-universal. In general it depends on the relative size of a wavelength and emitter—similar to optical blackbody radiation—but tends to a universal limit for macroscopic-sized emitters.

**4. Note on the experimental observation of the relaxation-induced sound**

The estimate according to the equation (12) indicates the emission of ~10 phonons per second. This seems to be within experimental limits for hearing of the "music of the spheres" emitted by the electronic bubbles in liquid helium if one can pump their 1P level in a quasi-continuous manner. However, I point at the disparity of this estimate and the naive estimate of the wasted energy corresponding to the inhomogeneous broadened transition. Namely, if the assumed bandwidth is on the order of 0.01 eV supported by the experiments [8, 15] then the Fourier-limited relaxation time is $10^{-11} \div 10^{-12}$ s$^{-1}$. Because of that, the relaxation of optical quantum with characteristic energy of 0.1 eV in the infrared requires the emission of $10^{10}$ -$10^{11}$ phonons per second into the liquid. I would be glad to ascribe nine-to-ten orders of magnitude difference between the estimate according to the equation (12) and the back-of-the envelope estimate above to the difference of the total phonon flux and the phonon flux in the far zone [6] of the acoustic radiation. However, current evidence is insufficient to support or reject this point of view.

## 5. Sound emission in the ground state?

We are accustomed to the fact that ground states of electronic systems do not radiate. This is clearly seen from the equation (12) because the power of radiated sound is proportional to the quadrupole matrix element squared, which becomes zero in the ground state of a bubble. However, the electronic ground state of the bubble is not necessarily its ground state in all other force fields acting on a quantum system, e.g. gravity. If the bubble acquires a constant speed due to buoyancy, the quasiparticles (phonons) reflected from the upper cap of the bubble are blue-shifted by acoustic Doppler effect. The quasiparticles reflected from the lower cap are similarly red-shifted. This indicates the drag acting on a bubble even in a superfluid, for finite temperatures. The energy lost to the drag must be converted into acoustic energy of phonons.

A crude estimate of this Doppler shift shows it to be sufficiently large to be distinct on the background of the assumed blackbody spectrum of phonons. The estimate reads as follows:

$$\Delta\omega_D \approx \omega \frac{V_b}{v_s} \approx \frac{v_s}{a} \times \frac{V_b}{v_s} \cong \frac{V_b}{a} \qquad (14)$$

In equation (14), $V_b$ is the buoyant velocity of the bubble and ω is a characteristic frequency of the scattered phonon. Natural buoyancy of the nanometer size bubble is too low to be discernible on the background of its Brownian motion but if the constant and uniform electric field is applied to the bubble, the upward mobility of the bubble can be observed with the existing experimental techniques [8, 15].

The power dissipated by the bubble moving in a superfluid can be estimated through the drag force obeying a cross between Stokes (proportionality of force to V and the lost energy to V²) and Newton (surface area scaling) laws:

$$N = -\xi \frac{\rho_n V_b^2}{2} S_b \qquad (15)$$

where $\rho_n$ is a density of a normal component of the liquid, $S_b$ is a bubble cross section and $\xi$ is a coefficient. The coefficient in equation (15) can be calculated by the methods similar to that of the Section 3.

Power losses by the bubble resulting from the recoil of phonons from a bubble wall can be expressed as the surface integral of the product of transferred energy $2\cdot\hbar\Delta\omega_D$, times the phonon flux $I_{ph} = v_s \cdot w(\mathbf{k})/4$, where $w(\mathbf{k})$ is the energy density of phonons at a given temperature:

$$N = -\iint \frac{\hbar \cdot \vec{V}_B \cdot \vec{k}}{2} \cdot v_s w(k) \cdot dS \cdot \frac{d^3k}{(2\pi)^3} \qquad (16)$$

Two integrals symbolically express integration over the components of the phonon wavevector **k**, and over the surface of the bubble.

The integral over the momentum of recoil would be equal to zero but for the Doppler shift of the reflected phonon. Assuming that the Doppler shift is much smaller than the frequency of the phonon, we can expand the denominator of the Equation (16) and then transfer a small exponential into the numerator. The first term in Taylor expansion of denominator, which contains Doppler shift in the first order, disappears on k-integration. The second term proportional to the Doppler shift in the second order in the factor $\eta = (\hbar \cdot V_B \cdot k/k_B \cdot T)$ provides a formula for the losses due to phonon drag in the liquid:

$$N = -\frac{4\hbar^2 \cdot v_s a^2}{\pi^2 k_B T} \int \frac{\exp(\hbar v_s k / k_B T) \cdot (\vec{V_B} \cdot \vec{k})^2}{\left(\exp(\hbar v_s k / k_B T) - 1\right)^2} d^3k \qquad (17)$$

Evaluation of the k-integral for the Equation (17) gives:

$$N = -\frac{\hbar \cdot v_s a^2}{6\pi^2} \left(\frac{k_B T}{v_s \hbar}\right)^4 I_3 V_B^2 \qquad (18)$$

where $I_3$ is the dimensionless integral:

$$I_3 = \int_{-\infty}^{\infty} \frac{e^x \cdot x^4 dx}{(e^x - 1)^2} = \frac{8\pi^4}{15} \cong 51.95 \qquad (19)$$

One observes a striking similarity of the expression of Equation (18) with the Landau formula for the phonon contribution into the density of the normal component of a superfluid. ([16], Chapter 2).

This, "Newtonian-Stokes" drag of a mesoscopic body in a fluid is intimately related to the fact that blackbody photon distribution has a wide spectrum. Thus, we can expect that a macroscopic body experiences drag in a Bose-Einstein condensate (BEC) in a trap because the excitation spectrum is always broadened because of the finite dimensions of a trap. On the contrary, this drag should disappear in an infinite condensate in full accordance with a notion of superfluidity. I provide an estimate of the BEC drag on a macroscopic sphere in Appendix B.

How this drag can be distinguished from other possible physical mechanisms of dissipation? One possibility would be to observe sound at a frequency corresponding to two-roton bound state $2 \cdot \Delta - E_B \cong \frac{2 \times 8.65\, K° \, k_B}{\hbar} - E_B$ despite the Boltzmann cutoff factor of $e^{-2\Delta - E_B / k_B T}$. [17] The sound at this frequency should disappear in the absence of bubbles moving through the fluid.

**Conclusion**

An electronic bubble ("bubblonium") moving through the superfluid dissipates energy through several channels. If the electronic state is excited, the energy is transferred to the liquid through quantum oscillations of a bubble. This oscillations diminish in intensity but do not disappear even at T=0 K°. Part of this energy is emitted in a form of high frequency ($10^{10}$ – $10^{11}$ Hz) sound. Despite high attenuation of sound of this frequency in liquid He, there is a hope that it can be observed using existing experimental techniques.

If the bubble moves through a liquid, it exhibits drag caused by uncompensated recoil from acoustic Doppler Effect of incoming phonons. In a very loose language one can imagine that a phonon reflecting from an upper cup of a bubble (we imagine the bubble buoyant) acquires energy while the phonon reflecting from a lower cup loses energy. The net effect of this reflections is not canceled in a second order of η—the ratio between phonon Doppler shift and thermal energy—and results in a drag on a bubble. This effect is obviously absent at T=0 K°. While a miniscule heating of liquid is hard to observe, nonlinear effects (binding of roton pairs) can tentatively be observed on the background of other excitations.

The diagrams for the phonon scattering by a surface of the bubble can be presented in the form, which caricatures Penrose diagrams for the Hawking radiation. At the present time, solid state analogues of astrophysical or particle physics phenomena are in much vogue among solid state researchers. [18, 19] Yet, in most of these solid state situations, they are poorly controllable with respect to the values of the parameters, for which analogues of the Hawking radiation are observed. Experimenters with bubblonium can vary many parameters: liquid ($H_2$, $He^3$, $He^4$), temperature, external fields, number of electrons inside a bubble so that significance of analogy can be raised to the level of modeling.

**Appendix A**. Estimates on the size of the quadrupole moment for the bubble.

In the paper Lerner-Chadwick-Sokolov [1], we derived a formula connecting standard deviation of a line shape for inhomogeneous broadening of a spectral line in bubblonium with the energy $E_{1P}$ for the 1p→1s transition. This formula reads as:

$$\delta\omega_{1p} \cong 0.616 \cdot \frac{E_{1P}}{\hbar}\sqrt{<\alpha_2^2>} \qquad (A.1)$$

For the assumed realistic value of $\delta\omega_{1p}$ = 0.01 eV, and the experimental value of the transition energy $E_{1P}$ = 0.182 eV (corresponding to the maximum absorption at $\lambda_{1P}$ = 6.8 µm), the value for the mean quadratic deformation of the bubble:

$$\sqrt{<\alpha_2^2>} \approx 8.9\%$$

i.e. quantum fluctuations deform the axes of the ellipsoid by approximately 9%. This is remarkable because the assumption of linear oscillations, which we made in [1] without giving it much thought, seem to hold beautifully. Mechanical deformation and the perturbation parameter in the Schrodinger equation are related by Equation [?] from [1], for historical reasons:

$$\sqrt{<\varepsilon_2^2>} = \sqrt{4/5 <\alpha_2^2>} \qquad (A.2)$$

i.e. the numerical value to be used for the estimation Equation (12) is equal to 8%. It weakly (as a square root) depends on the fraction of the linewidth attributable to inhomogeneous broadening.

**Appendix B.** The drag on macroscopic body in BEC.

To compute the drag, we need to return to the Equation (16). Here, unlike the thermal distribution of quasi-particles, we assume Gaussian distribution in the symmetric trap

$$w(k) = \frac{n_p \hbar^3}{(2\pi)^{3/2} \sigma_p^3} e^{-\frac{(\hbar k)^2}{2\sigma_p^2}} \tag{B.1}$$

where $\sigma_p$ is the momentum uncertainty of a trapped particle and $n_p$ is a particle density. By an order of magnitude $\sigma_p$ is equal to

$$\sigma_p = \frac{\hbar}{\sqrt{<x^2>}} \tag{B.2}$$

where <x> is an average localization length of a condensate particle in a trap.

Calculations, which are absolutely similar to those in the Section 5, lead to the following expression:

$$N = -\frac{8 n_p \cdot a^2 \sigma_p}{3 \cdot (2\pi)^{1/2}} V_B^2 \tag{B.3}$$

The only difference with the blackbody case is that the number of quasi-particles is fixed, unlike the blackbody case where it is determined by the same temperature *T* as the momentum spread of quasiparticle distribution.

One sees (1) that the drag disappears when the dispersion of the quasi-particles disappears (for instance, in the infinite condensate) and (2) disappears for $\hbar \to 0$ or infinitely massive particle of a condensate because of the De Broglie-type formula (B.2). These facts indicate purely quantum character of the expression of Equation (B.3). The force acting upon a macroscopic scatterer can be too weak to measure. However, force acting upon condensate can be possible to measure in the configuration sketched on the Fig. 2. The sketch substantially depends on the fact that once released from a trap, a condensate generally retains momentum uncertainty of its wavepacket.

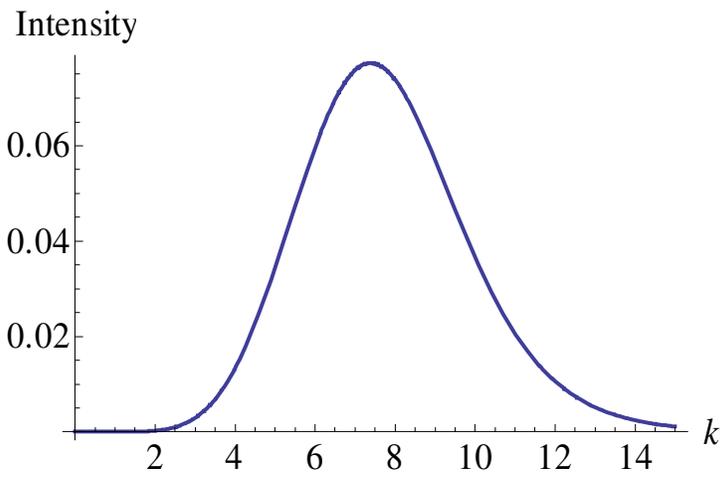

a)

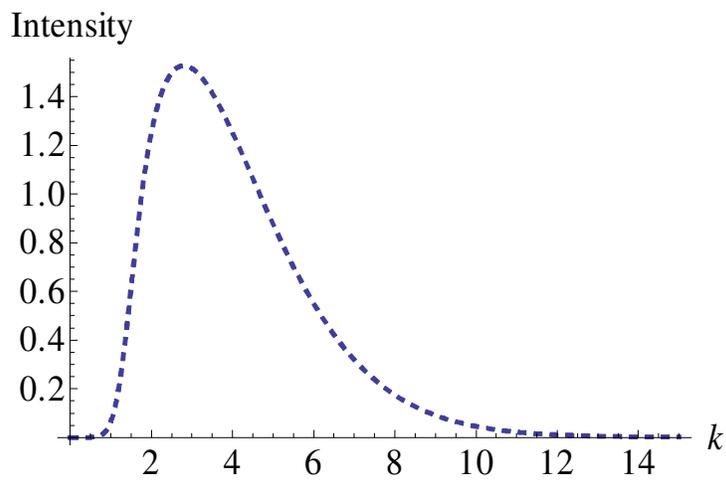

b)

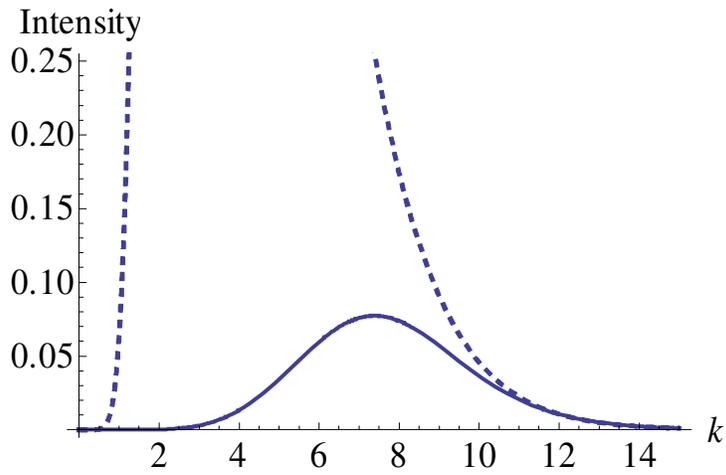

c)

Fig. 1 Intensity spectrum of sound emission by the electronic bubble in helium as a function of dimensionless wavevector $k=q\cdot a$. a) Parameter $x = \frac{\hbar v_s q}{k_B T}=0.25$, b) $x=1.5$, c) both curves are plotted to the same scale. For the large bubble, the law becomes universal, asymptotically coinciding with the law for blackbody radiation.

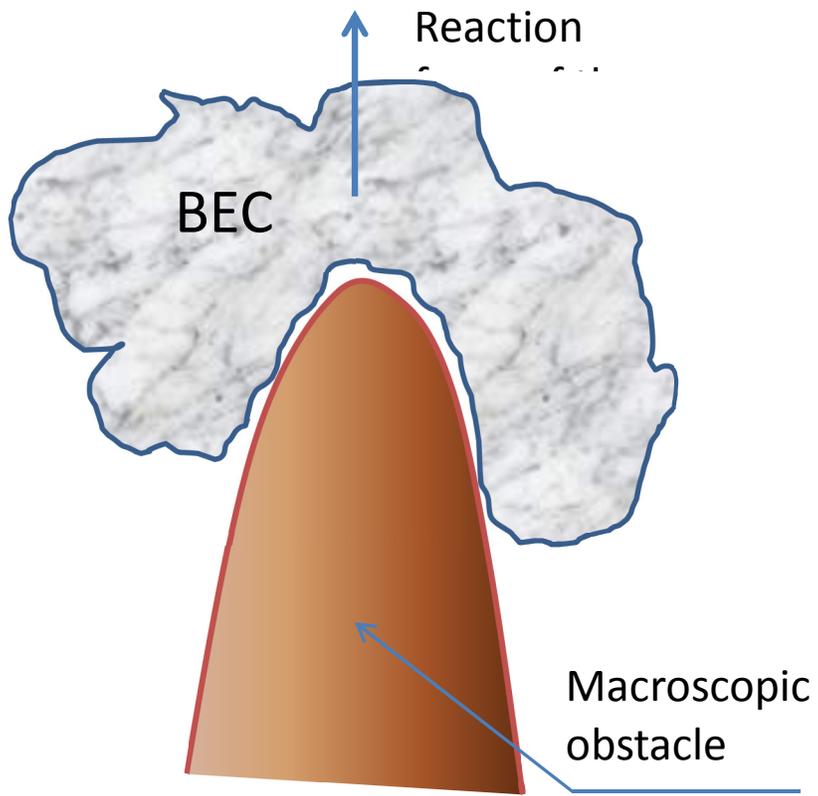

Fig. 2 Sketch of an experiment for observation of the stopping force experienced by the Bose-Einstein Condensate (BEC) due to the phonon drag.